# Current control of spin-wave transport in YIG via the $VO_2$ insulator-to-metal transition

Aleksei A. Nikitin[1*], Nikolai Kuznetsov[1], Lukáš Flajšman[1], Erkki Lähderanta[2,3], Sebastiaan van Dijken[1*]

[1]NanoSpin, Department of Applied Physics, Aalto University School of Science, P.O. Box 15100, FI-00076 Aalto, Finland
[2]Department of Physics, LUT University, 53850 Lappeenranta, Finland
[3]Department of Physics, School of Science, Politecnico di Milano, Como 22100, Italy
To whom correspondence should be addressed; E-mail: aleksei.a.nikitin@gmail.com; sebastiaan.van.dijken@aalto.fi

*Abstract*—**Electrically reconfigurable spin-wave transport is central to realizing energy-efficient magnonic information technologies, yet existing current-driven approaches require current densities that undermine the fundamental dissipation advantage of spin-wave-based signal processing. Here, we demonstrate low-power electrical control of spin-wave propagation in a hybrid magnonic device comprising a low-damping yttrium iron garnet (YIG) film coupled to an adjacent vanadium dioxide ($VO_2$) stripe. Passing a small current through $VO_2$ triggers its insulator-to-metal transition (IMT), producing a sharp, reversible resistivity reduction exceeding three orders of magnitude. This abrupt phase transition locally modifies the spin-wave dispersion in YIG, reducing the group velocity and suppressing transmission by up to 40 dB near the ferromagnetic resonance frequency. Pulsed current excitation enables reconfiguration of spin-wave transport on a 10 ms timescale. Temperature-dependent measurements, corroborated by thermal modeling, decouple the respective contributions of current-induced heating of YIG and the IMT in $VO_2$, establishing the latter as the dominant control mechanism. Our results demonstrate that electrically induced phase transitions in correlated oxides constitute an energy-efficient strategy for dynamic spin-wave manipulation, providing a scalable route toward reconfigurable magnonic devices.**

## I. Introduction

Energy-efficient control of propagating spin waves, or magnons, is essential for next-generation information processing technologies that rely on active manipulation of spin-wave interference or coupling (*1–3*). Unlike conventional charge-based electronics, magnonic systems encode information in the amplitude or phase of a spin wave and transport it via angular momentum transfer rather than electrical current, thereby significantly reducing energy consumption by eliminating Joule heating. A wide variety of magnonic computing elements has already been demonstrated, including spin-wave logic gates (*1, 4–6*), three-terminal transistors (*7–9*), directional couplers (*10*), oscillators and resonators (*11-14*), and neuromorphic and other unconventional architectures based on spin-wave transport (*15–19*). Active control of spin-wave amplitude or phase in these devices can be achieved through coupling to electric currents (*20–28*), applied voltages (*29–32*), or optical excitation (*33–35*). Among these approaches, electrical control via current or voltage is particularly attractive owing to its compatibility with device miniaturization and large-scale integration. However, current-based interactions with spin waves, including Oersted field generation (*20–22*), Joule heating (*23*), spin-orbit torques (SOTs) (*24–27*), or current-driven domain rotation (*28*), often require substantial electrical power, potentially offsetting the intrinsic efficiency advantage of low-energy spin-wave transport. The current densities required for spin-wave manipulation vary widely: domain rotation can occur at relatively low values (~$10^5$ A $cm^{-2}$), whereas Joule heating and SOT-driven mechanisms typically require $10^6$ – $10^8$ A $cm^{-2}$, and Oersted-field approaches demand increasingly high densities as device dimensions shrink. Voltage-controlled magnonic structures offer improved energy efficiency (*29–32*) but face their own challenges, including the inherently weak influence of electric fields on magnetism and limitations related to materials growth, fabrication, and device integration.

Interfacing a low-damping magnetic insulator such as yttrium iron garnet (YIG) with a nonmagnetic conductor can strongly modify the spin-wave dispersion (*36, 37*). The magnitude of this effect depends sensitively on the electrical resistivity of the nonmagnetic layer, enabling electrical control of spin-wave transport through reversible resistivity changes. Vanadium dioxide ($VO_2$) is a particularly promising material for this purpose, as it undergoes a well-studied insulator-to-metal transition (IMT) slightly above room temperature (*38–41*). Theoretical studies of YIG/$VO_2$ heterostructures have predicted pronounced changes in spin-wave dispersion across this transition (*42, 43*), and experimental validation, achieved by heating the sample through the $VO_2$ phase transition using a temperature-controlled stage, has confirmed these predictions (*44*), revealed tunable spin-charge conversion (*45*), and demonstrated modulation of SOTs (*46*). However, this approach is slow and requires heating the entire sample, including the YIG layer, rendering it unsuitable for practical magnonic applications.

Here, we demonstrate efficient electrical control of spin-wave transport in a YIG film via a current applied to an adjacent $VO_2$ stripe. The current induces two concurrent effects. First, it triggers the IMT in $VO_2$ at a threshold current density of approximately $J = 1.0 \times 10^3$ A cm$^{-2}$, causing a sharp resistivity reduction that locally modifies the spin-wave dispersion and reduces the group velocity in YIG, suppressing transmission by 25 dB at $J = 1.4 \times 10^4$ A cm$^{-2}$. Second, modest heating of the YIG film near the excitation antenna shifts the spin-wave transmission spectrum toward lower frequencies. Acting together, these two mechanisms produce a combined transmission suppression of approximately 40 dB near the ferromagnetic resonance (FMR) frequency of YIG. Time-resolved measurements demonstrate that the modulation is fully reversible on a 10 ms timescale, consistent with the intrinsic switching dynamics of the $VO_2$ phase transition.

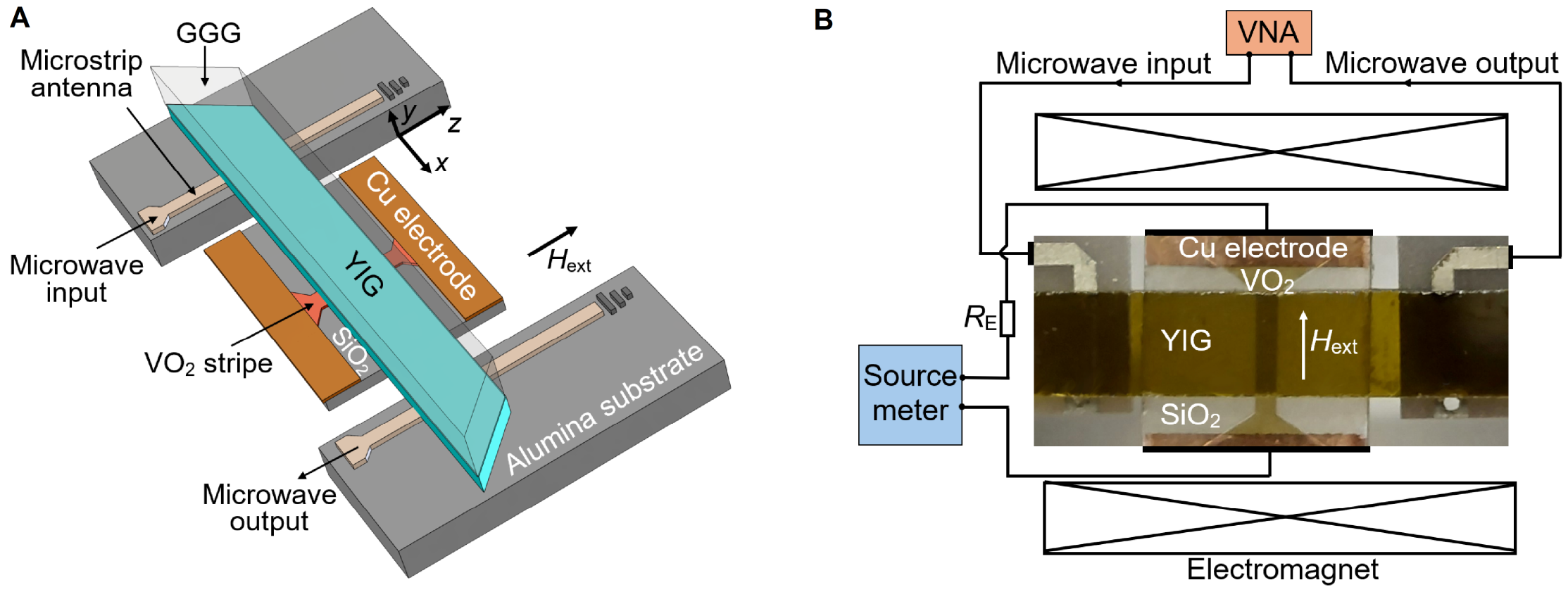

**Fig. 1. Current-controlled magnonic device. (A)** Schematic illustration of the device architecture, comprising a YIG film, a $VO_2$ stripe, and two parallel microstrip antennas. **(B)** Optical micrograph of the fabricated device and schematic of the measurement geometry used for DC current control. Spin waves are excited and detected by the microstrip antennas via a VNA, while the resistivity of the $VO_2$ stripe is tuned by a DC current supplied by a source meter.

## II. Results

### A. *Magnonic device and measurement geometry*

The magnonic device consists of a 5.7-µm-thick single-crystal YIG film grown by liquid-phase epitaxy on a (111)-oriented gadolinium gallium garnet (GGG) substrate (30 × 2 mm$^2$), a 650-nm-thick, 300-µm-wide $VO_2$ stripe deposited on an amorphous quartz substrate by DC magnetron sputtering and contacted with Cu electrodes, and two parallel 50-Ω microstrip antennas connected to a two-port vector network analyzer (VNA). As shown in Fig. 1, the YIG film is placed face-down onto the microstrip antennas, which are separated by 5 mm, with a single $VO_2$ stripe positioned between them in direct contact with the YIG surface. Spin-wave propagation in the YIG film is characterized by recording the $S_{11}$ and $S_{21}$ scattering parameters. Active control of spin-wave transmission is achieved by passing an electrical current through the $VO_2$ stripe using a source meter. All DC measurements are conducted in constant-current mode with a 250-Ω current-limiting resistor connected in series. Time-resolved measurements of the $S_{21}$ parameter at fixed frequency are carried out by incorporating a MOSFET switchboard into the measurement circuit. In all experiments, an in-plane magnetic bias field applied parallel to the antennas establishes the Damon-Eshbach transport geometry ($k \perp H$). Further

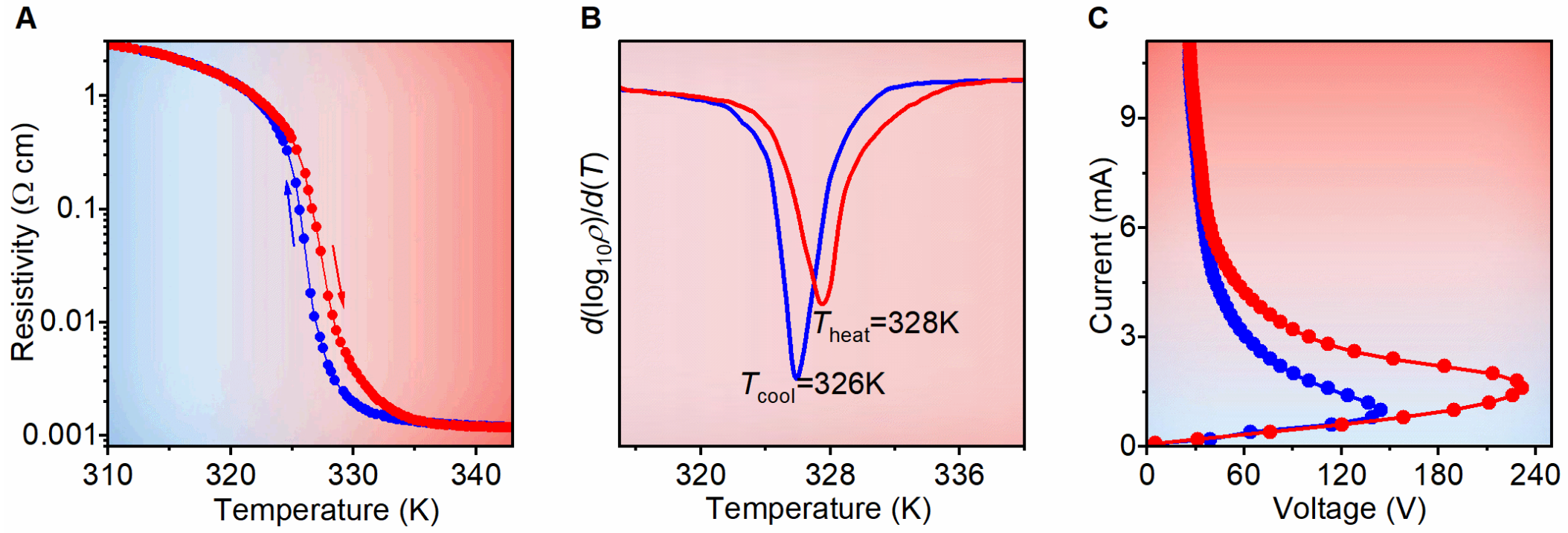

**Fig. 2. Temperature- and current-driven IMT in $VO_2$.** (**A**) Resistivity ($\rho$) of the $VO_2$ stripe as a function of temperature, measured during heating and cooling cycles. (**B**) Temperature derivative $d(\log_{10}\rho)/dT$ as a function of temperature; the peak positions of the heating and cooling curves reveal a hysteresis of approximately 2 K. (**C**) Voltage across the $VO_2$ stripe as a function of applied current at 298 K, demonstrating a reversible, current-driven IMT, evident as an abrupt drop in voltage above a threshold current. Red and blue curves correspond to increasing and decreasing current, respectively.

experimental details are provided in the Materials and Methods section.

## B. *Temperature and current control of the IMT in $VO_2$*

Before investigating active spin-wave manipulation, we first characterize the IMT in the $VO_2$ stripe. Figure 2A shows the resistivity response of $VO_2$ as its temperature is cycled through the transition region. The resistivity changes by more than three orders of magnitude within a 30 K window, with a hysteresis of 2 K between the heating and cooling curves (Fig. 2B), consistent with the monoclinic-to-tetragonal structural transition in high-quality $VO_2$ crystals (*38–41*). In addition to thermal control, the resistivity can be tuned electrically at room temperature by applying a current to the $VO_2$ stripe (Fig. 2C). The current-induced transition from the high- to low-resistivity state commences at approximately 2 mA, corresponding to a low current density of $1.0 \times 10^3$ A cm$^{-2}$. Reducing the current back to zero fully restores the insulating state.

## C. *Current control of spin-wave transport*

The influence of passing a DC current through the $VO_2$ stripe on spin-wave transport in the YIG film was characterized by VNA measurements of the return loss ($S_{11}$) and insertion loss ($S_{21}$) as a function of frequency under applied magnetic fields ranging from 0 to 27 mA (current densities $J = 0 - 1.4 \times 10^4$ A cm$^{-2}$). Figure 3A shows the real part of the $S_{11}$ spectra at three representative current densities under a bias field of 144 mT. The onset of increased return loss identifies the YIG FMR frequency near the excitation antenna. With increasing DC current, the FMR frequency shifts to lower values, a result of moderate Joule heating in the $VO_2$ stripe and subsequent heat diffusion into the YIG film (discussed below). At finite current, the $S_{11}$ spectra exhibit oscillations above the FMR frequency, characteristic of propagating spin waves (*47, 48*) that are partially reflected by the $VO_2$ stripe back toward the excitation antenna (*49*).

The transmitted spin-wave signal ($S_{21}$; Fig. 3B) reveals two concurrent effects with increasing current density: a downshift in frequency and a pronounced attenuation of the transmitted signal. The frequency downshift mirrors the behavior observed in $S_{11}$ and is attributed to a small increase in YIG temperature near the excitation antenna. The attenuation, by contrast, reflects current-induced suppression of spin-wave transport in the vicinity of the $VO_2$ stripe, with a marked reduction in transmission emerging above $J > 1.0 \times 10^3$ A cm$^{-2}$, as shown in Fig. 3C. Three mechanisms may underlie this suppression: (i) the Oersted field generated by the DC current, (ii) heating of the YIG film, and (iii) the IMT in $VO_2$. The Oersted field contribution can be ruled out: identical $S_{21}$ spectra are obtained for opposite current polarities, and the calculated Oersted field at the center of the YIG film does not exceed 0.016 mT at maximum current density (Supplementary Materials), far below the threshold of any measurable effect.

To decouple thermal and resistive contributions, measurements were caried out with the sample mounted on a temperature-controlled stage. Figure 4 compares data acquired at 298 K and 282 K under $J = 1.1 \times 10^4$ A cm$^{-2}$, alongside simulated temperature profiles (see Materials and Methods). At 298 K, Joule heating raises the $VO_2$ temperature to 335 K, surpassing the IMT threshold and driving $VO_2$ into the metallic state. At 282 K, the $VO_2$ temperature reaches only 317 K, remaining below the transition and preserving the insulating phase. As a result, the $VO_2$ resistivity differs by approximately three orders of magnitude between these two conditions (Fig. 2). In both cases, the heated $VO_2$ stripe induces a comparable temperature gradient in the YIG film. At 298 K, the YIG temperature rises from ~302 K near the microstrip antennas to ~335 K in the $VO_2$ contact region ($\Delta T \approx$ 33 K). At 282 K, the corresponding range is 285 – 317 K ($\Delta T \approx$ 32 K). The heating primarily reduces the saturation magnetization $M_s$ (*34*), which varies approximately linearly with temperature in this range, yielding a nearly identical relative change of –1.4% (*33*) in both cases. Since the thermal profiles in YIG are essentially equivalent under the two-base-temperature conditions, local heating near the $VO_2$ stripe cannot account for the transmission suppression observed at 298 K but absent at 282 K. This behavior must therefore be attributed almost exclusively to the large difference in $VO_2$ resistivity. The downward shift of the transmission spectra observed in both cases reflects a modest increase (~3 – 4 K) in YIG temperature near the excitation antenna under the applied current. This interpretation is corroborated by two independent lines of evidence. First, the current density required to trigger the $VO_2$ phase transition (Fig. 2) closely coincides with the onset of transmission suppression (Fig. 3C). Second, using the theoretical framework of Ref. *43*, we modeled spin-wave transmission across the YIG film in the presence of the $VO_2$ IMT (Supplementary Materials). Combined with modest heating of YIG near the excitation antenna, the model quantitatively reproduces the measured $S_{21}$ spectra and the magnitude of transmission suppression (green dashed lines in Fig. 3B, C).

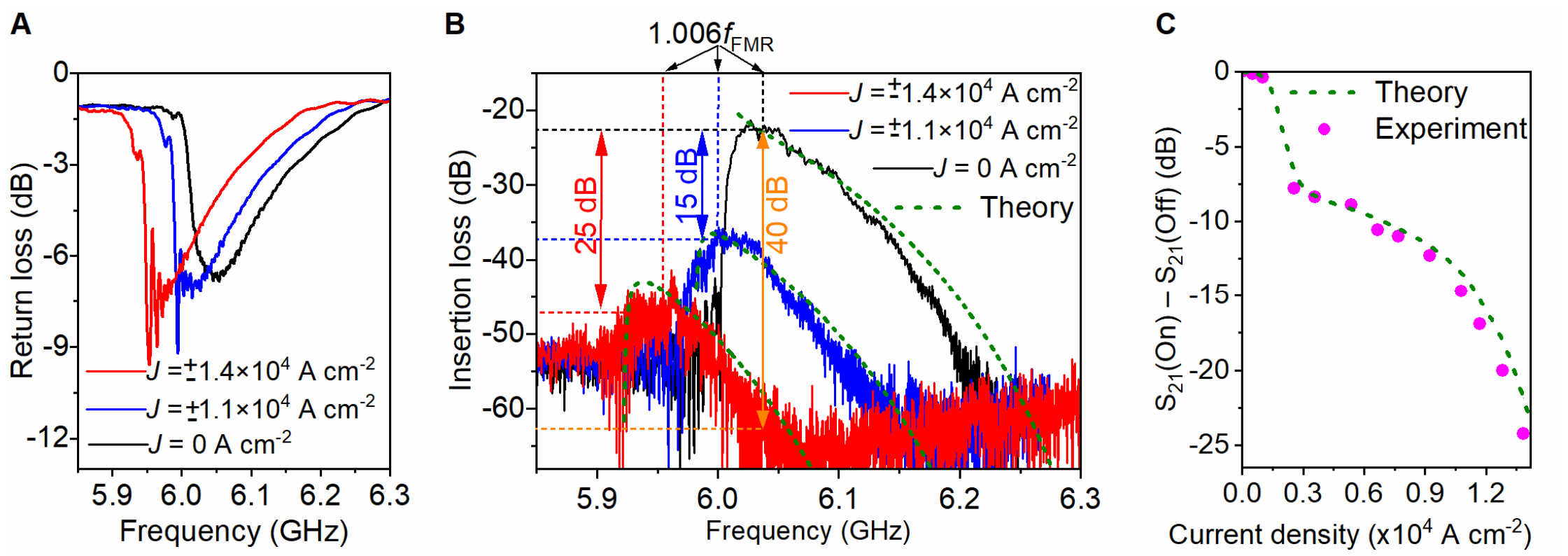

**Fig. 3. Current control of spin-wave transport.** (**A** and **B**) $S_{11}$ and $S_{21}$ magnitude spectra measured at representative DC current densities. In (B), vertical dashed lines mark the frequencies $1.006f_{FMR}$ corresponding to each applied current density. Suppression of spin-wave transmission is indicated by blue and red double-headed arrows for $J = 1.1 \times 10^4$ A cm$^{-2}$ and $J = 1.4 \times 10^4$ A cm$^{-2}$, respectively. The additional signal attenuation at a fixed frequency of 6.04 GHz for $J = 1.4 \times 10^4$ A cm$^{-2}$ is indicated by an orange double-headed arrow. All spectra were acquired at 298 K under an applied magnetic field of 144 mT. Green dashed lines in (B) represent theoretical calculations of the $S_{21}$ spectra (see Supplementary Materials). (**C**) Suppression of the spin-wave signal at $1.006f_{FMR}$ as a function of current density, isolating the IMT-induced attenuation from the concurrent frequency downshift. Symbols epresent experimental data, and the green dashed line shows the corresponding theoretical calculation.

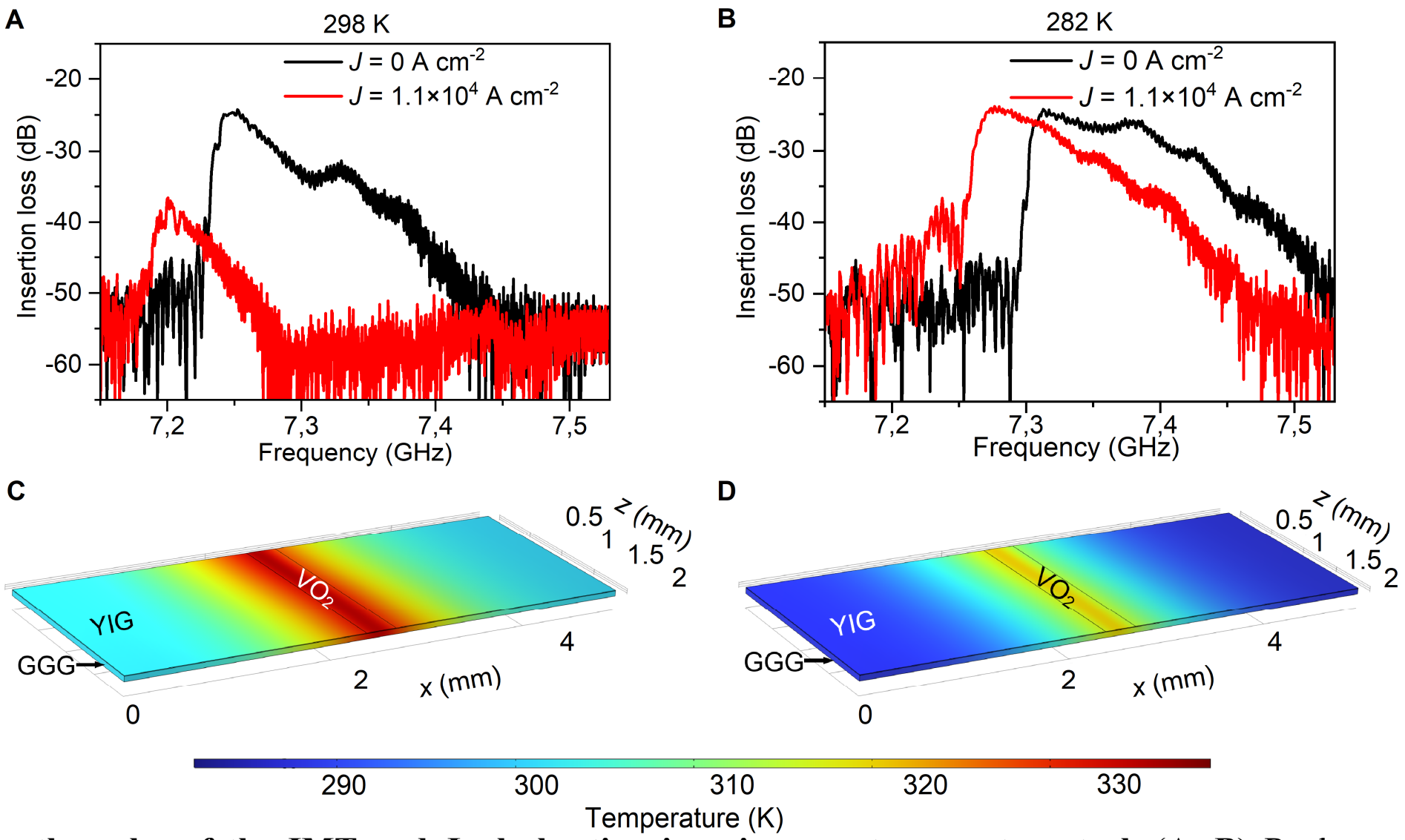

**Fig. 4. Disentangling the roles of the IMT and Joule heating in spin-wave transport control.** **(A, B)** Real parts of the $S_{21}$ spectra measured at an applied magnetic field of 185 mT for $J = 0$ and $J = 1.1 \times 10^4$ A cm$^{-2}$ at device temperatures of 298 K and 282 K, respectively. A pronounced suppression of spin-wave transmission is observed in (A) but absent in (B). **(C, D)** Simulated temperature distributions in the GGG substrate, YIG film, and $VO_2$ stripe for $J = 1.1 \times 10^4$ A cm$^{-2}$ at initial temperatures of 298 K and 282 K. At 298 K, Joule heating raises the $VO_2$ stripe temperature to 335 K, driving it above the IMT threshold and into the metallic state. At 282 K, the stripe reaches only 317 K, remaining in the insulating phase. Based on Fig. 2A, these conditions correspond to an approximately three-order-of-magnitude difference in $VO_2$ resistivity. The comparable temperature increase in the YIG film under both conditions demonstrates that local Joule heating cannot account for the strong suppression of spin-wave transmission observed in (A); pronounced attenuation occurs only when the $VO_2$ stripe is driven into its metallic state. The downshift of the FMR frequency observed at both temperatures is attributed to a moderate increase in YIG temperature near the microstrip antennas under the applied current.

The degree of suppression is quantified just above the FMR frequency at $1.006f_{FMR}$ (vertical dashed lines in Fig. 3B). At $J = 1.1 \times 10^4$ A cm$^{-2}$, the spin-wave signal is attenuated by 15 dB, increasing to 22 dB at $J = 1.4 \times 10^4$ A cm$^{-2}$. When evaluated at a fixed frequency (6.04 GHz, corresponding to $1.006f_{FMR}$ at zero current), the combined effect of attenuation and frequency shift yields a suppression of up to 40 dB. Notably, the current densities required to achieve this level of control are substantially lower than in other current-driven approaches (*20–28*), underscoring the potential of this mechanism for energy-efficient spin-wave manipulation.

To further characterize the current-controlled magnonic device, we measured the frequency dependence of the group delay time $\tau_g$ using the VNA. Figure 5A shows results for three current densities, together with the corresponding spin-wave group velocity ($v_g = l/\tau_g$, where $l$ is the antenna separation). At zero current, $\tau_g$ exhibits a nonlinear frequency dependence reflecting the intrinsic dispersion of Damon-Eshbach spin waves in YIG. Upon applying current to the $VO_2$ stripe, $\tau_g$ increases (i.e., $v_g$ decreases) and the transmission bandwidth narrows. Both effects originate from the flatting of the spin-wave dispersion as $VO_2$ undergoes the IMT (Supplementary Materials). At $1.006f_{FMR}$, $\tau_g$ increases by approximately 60% at the highest applied current. The ability to tune $\tau_g$ efficiently with relatively modest currents is particularly attractive for magnonic reservoir computing schemes based on time-multiplexed nonlinear delay lines (*16, 17*).

Figure 5B presents the magnetic-field dependence of current-controlled spin-wave transport. Increasing the applied field shifts the spin-wave frequencies to higher values and simultaneously enhances the current-induced suppression of the spin-wave signal. The stronger attenuation at higher fields is attributed to a more pronounced increase of the spatial damping decrement within the region of the YIG film contacted by the $VO_2$ stripe. Further analysis and supporting data are provided in the Supplementary Materials.

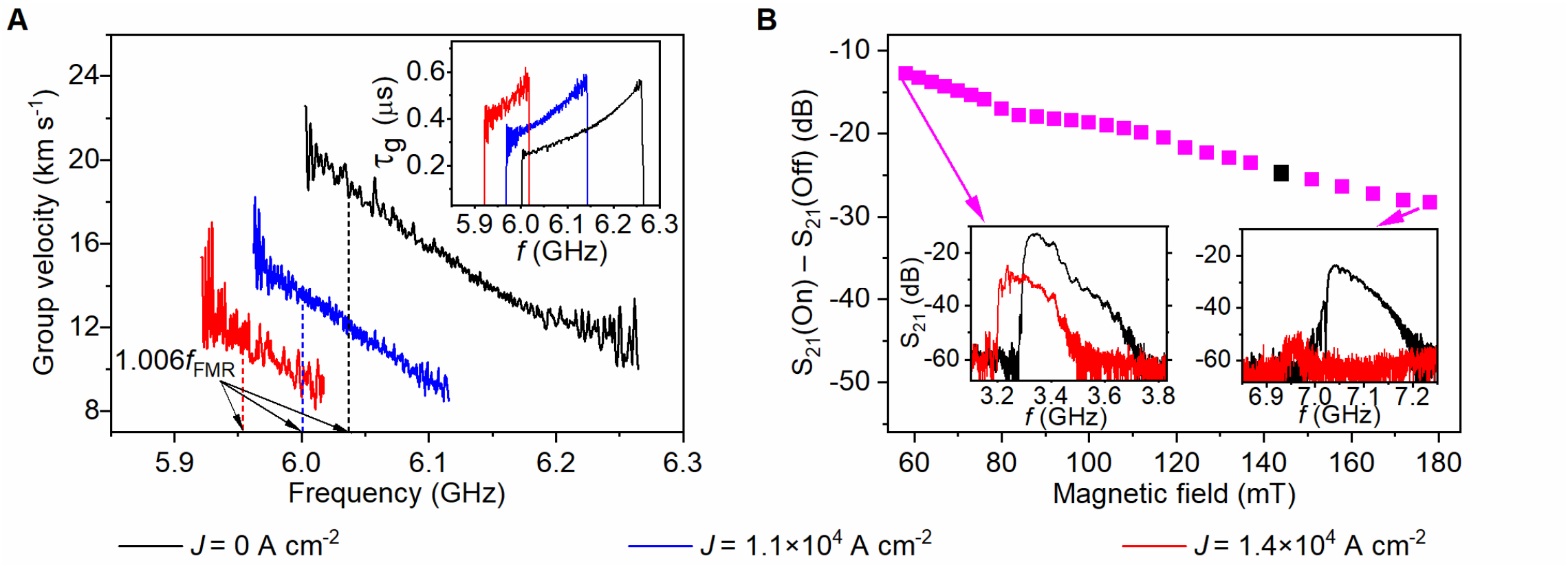

**Fig. 5. Group delay, group velocity, and magnetic-field dependence of spin-wave transport.** (**A**) Group delay time $\tau_g$ and corresponding group velocity $v_g$ as functions of frequency for applied current densities of $J$ = 0, 1.1 × 10$^4$, and 1.4 × 10$^4$ A cm$^{-2}$. Vertical dashed lines indicate the frequencies 1.006$f_{FMR}$ for each applied current density. (**B**) Magnetic-field dependence of spin-wave suppression at 1.006$f_{FMR}$ for $J$ = 1.4 × 10$^4$ A cm$^{-2}$. Insets show representative $S_{21}$ spectra acquired at low and high applied magnetic fields. The black square marks the condition (144 mT, $J$ = 1.4 × 10$^4$ A cm$^{-2}$) also shown in Fig. 3.

D. *Electrical control of spin-wave transport using current pulses*

To investigate the effect of short current pulses on spin-wave transport, we performed time-resolved measurements of the $S_{21}$ parameter at a fixed excitation frequency. Current pulses were injected into the $VO_2$ stripe and the 250-Ω series resistor via a voltage-controlled MOSFET switchboard integrated into the measurement circuit. The pulse duration was defined by the control voltage pulses (3.3 V), while the current amplitude was set by the source meter (see Methods). The sample temperature was maintained at 318 K to reduce the threshold current for the $VO_2$ IMT and to ensure stable switching-circuit operation, as discussed in the Supplementary Materials.

Figure 6A shows the temporal response of the spin-wave transmission at an excitation frequency of 7.49 GHz under a current-pulse train with a repetition rate of 10 Hz and a pulse duration $\tau$ = 20 ms at a bias field of 196 mT. Current pulses at $J$ = 0.25 × 10$^3$ A cm$^{-2}$ produce no measurable modulation of the transmission signal, indicating that the applied current is insufficient to drive the $VO_2$ stripe out of its insulating state. Increasing the current density to $J$ = 1.0 × 10$^3$ A cm$^{-2}$ results in a repeated suppression of the spin-wave signal by approximately 10 dB during each pulse. This suppression does not increase significantly upon further raising the current density to $J$ = 1.5 × 10$^3$ A cm$^{-2}$, suggesting that the IMT is already largely complete. The spin-wave response to individual current pulses exhibits noticeable variability, which we attribute to the stochastic nature of resistivity switching in $VO_2$ (*50, 51*). Specifically, the IMT proceeds via nucleation and growth of metallic domains within the insulating matrix, leading to spatial and temporal fluctuations in the formation of conductive pathways that are directly reflected in the spin-wave transmission signal.

The effect of pulse duration on spin-wave suppression is shown in Fig. 6B for $J$ = 1.0 × 10$^3$ A cm$^{-2}$. The transmission initially decreases over a characteristic timescale of 6 ms, largely independent of pulse duration. For longer pulses, this initial suppression is followed by a partial recovery before a subsequent decrease occurs. Upon termination of the current pulse, the spin-wave transmission rapidly returns to its pre-pulse level, confirming a reversible and non-destructive switching process. The non-monotonic temporal evolution reflects the anomalous current-voltage characteristics of $VO_2$ near the IMT (see Supplementary Materials and Refs. *52–54*). In this regime, Joule heating drives a percolative transition through the nucleation and growth of metallic domains: the initial 6 ms resistivity suppression corresponds to the rapid formation and expansion of conductive pathways, while the subsequent transient increase in resistivity is associated with the redistribution of current density and thermal gradients, which can temporarily disrupt percolation pathways. At longer times, the system evolves toward a more homogeneous metallic state. From the initial suppression timescale and the $VO_2$ stripe length of 4 mm, we estimate a phase-front propagation velocity of approximately 0.7 nm $ns^{-1}$, in reasonable agreement with recent studies on comparable $VO_2$ structures (*51*).

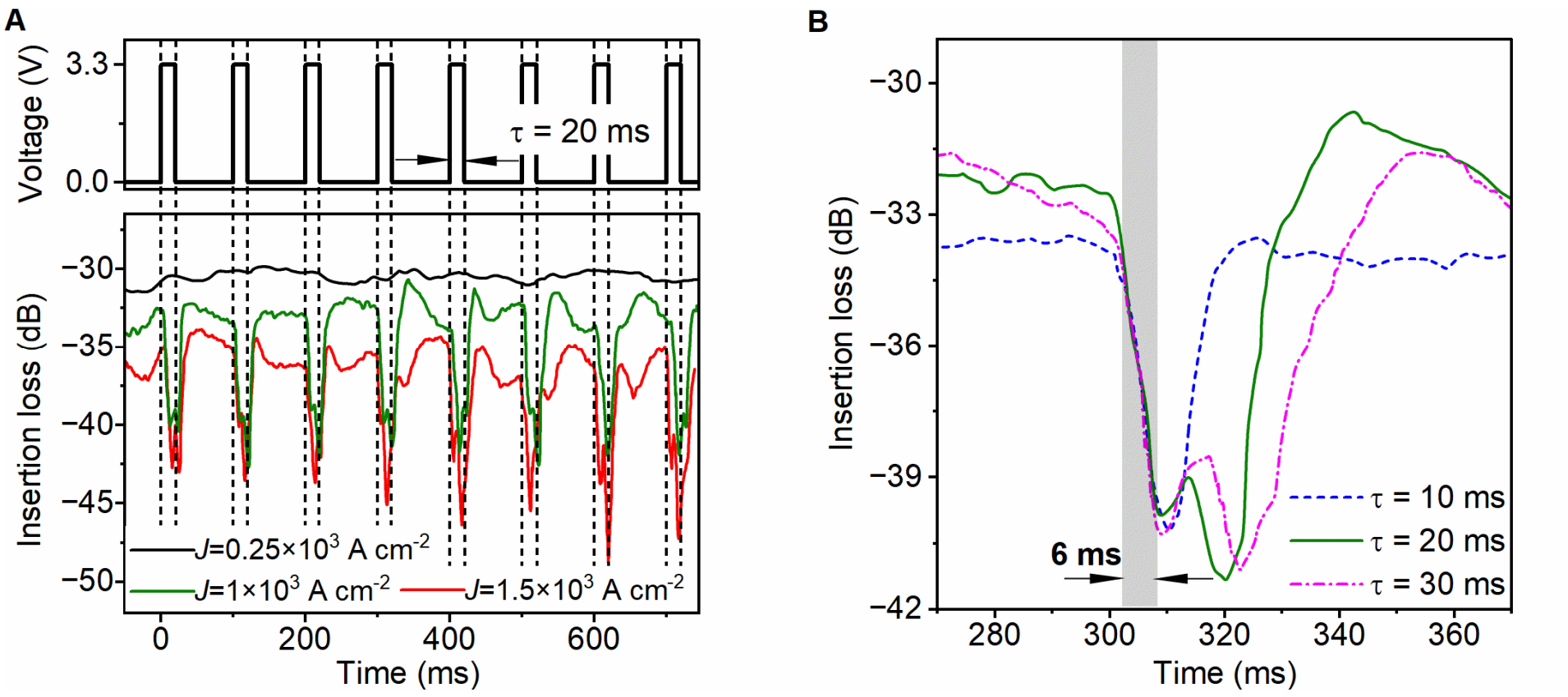

**Fig. 6. Time-resolved measurements of spin-wave transmission control.** (**A**) Temporal response of the $S_{21}$ signal to a current-pulse train with a repetition rate of 10 Hz and a pulse duration $\tau$ = 20 ms, measured at an excitation frequency of 7.49 GHz and an applied magnetic field of 196 mT, for current densities $J$ = 0.25 × 10$^3$, 1.0 × 10$^3$, and 1.5 × 10$^3$ A cm$^{-2}$. Results are shown for three representative current densities. The top panel shows the control voltage applied to the MOSFET switchboard that turns the current on and off. (**B**) Temporal response of the $S_{21}$ signal to single current pulses of duration $\tau$ = 10, 20, and 30 ms at a current density $J$ = 1.0 × 10$^3$ A cm$^{-2}$. The initial suppression over the first 6 ms is attributed to the rapid formation and expansion of conductive pathways within the $VO_2$ stripe. All measurements were performed at $T$ = 318 K.

## III. Discussion

Our results establish the IMT in $VO_2$ as an effective and energy-efficient mechanism for electrically controlled spin-wave transport. The energy cost of this approach is directly linked to the current density required to trigger the IMT, which decreases as the operating temperature approaches the transition temperature. This temperature dependence is clearly evidenced by comparing the pulsed experiments in Fig. 6, conducted at 318 K close to the IMT, with the DC measurements performed at 298 K (Figs. 3–5): operation near the transition temperature requires substantially lower current densities. For room-temperature applications, reducing the IMT temperature by 20-30 K would therefore directly lower the operating current density and further improve energy efficiency. This can in principle be achieved through elemental doping (*55, 56*), epitaxial strain engineering (*57, 58*), or stoichiometry control of $VO_2$ thin films (*59, 60*), all of which have been shown to shift the transition temperature by tens of kelvin without substantially degrading the resistivity contrast across the IMT.

A complementary strategy for reducing the operating current density is geometric downscaling of the $VO_2$ stripe. The power required to heat the active volume to the IMT temperature scales with the volume of the material; reducing the stripe dimensions therefore directly lowers the drive current density needed to trigger the phase transition. Downscaling offers two additional benefits: it reduces the IMT switching time, and the lower thermal mass of the active region accelerates heat dissipation once the current pulse is removed, thereby shortening the recovery timescale. In $VO_2$ thin-film devices with channel lengths ranging from tens of nanometers to a few micrometers, both switching and recovery times on the nanosecond timescale have been demonstrated (*61, 62*), an improvement of several orders of magnitude relative to the ~10 ms timescale observed in the macroscopic $VO_2$ stripes employed in the present study.

Downscaling the magnonic device geometry reduces the degree of spin-wave transmission suppression. In particular, narrowing the $VO_2$ stripe width shortens the propagation distance over which the spin-wave dispersion is modified by the resistivity change accompanying the IMT. To quantify this effect, we employed our theoretical model, which accurately reproduces the experimental transmission spectra. The results are summarized in Supplementary Fig. S9A. As the $VO_2$ stripe width is reduced, the IMT-induced suppression of the spin-wave signal decreases linearly, in excellent agreement with experiments performed on devices with $VO_2$ stripe widths of 200, 300, and 500 μm. The YIG film thickness similarly affects spin-wave control, though the dependence is more complex: in general, reducing the film thickness also lowers the degree of signal suppression. For example, decreasing the YIG thickness from 5.7 μm to 200 nm (300-μm-wide $VO_2$ stripe) reduces the spin-wave transmission suppression from 15 dB to 11 dB at a current density of $J = 1.1 \times 10^4$ A $cm^{-2}$ at $T = 298$ K (Supplementary Fig. S9B).

In summary, the YIG/$VO_2$ heterostructure introduced here represents a distinct and practically viable approach to reconfigurable magnonics, exploiting the abrupt, electrically driven phase transition of a correlated oxide to locally engineer the spin-wave dispersion without requiring large drive currents. With a threshold current density of $J = 1.0 \times 10^3$ A $cm^{-2}$ at room temperature, two to five orders of magnitude below that of other current-driven schemes, transmission suppression of up to 40 dB, and full reversibility on the 10 ms timescale, this platform ranks among the most energy-efficient electrically reconfigurable magnonic devices reported to date. The decoupling of thermal and resistive contributions established here clarifies the dominant control mechanism and provides a firm physical basis for further device optimization. Geometric downscaling of the $VO_2$ element and targeted reduction of the IMT temperature through doping or strain engineering offer concrete, materials-compatible pathways to faster switching speeds and room-temperature operation at reduced power. Beyond spin-wave modulation, the ability to gate magnonic transport via a correlated-oxide phase transition may find broader application in magnonic reservoir computing, reconfigurable delay lines, and hybrid magnonic-electronic circuits, establishing electrically induced phase transitions as a general design principle for low-dissipation magnonic information processing.

## IV. Materials and methods

### A. *Device fabrication*

The 5.7-μm-thick YIG film was grown on a (111)-oriented GGG substrate by liquid-phase epitaxy. A 30 × 2 $mm^2$ spin-wave waveguide was defined by dicing the YIG/GGG sample with a diamond wire saw. To minimize reflections of propagating spin waves, both ends of the waveguide were cut at 45°. The YIG film exhibited an FMR linewidth of $\Delta H = 31.83$ A $m^{-1}$ at 6 GHz and a saturation magnetization of $M_s = 1.39 \times 10^5$ A $m^{-1}$, consistent with bulk YIG values.

Spin-waves were excited and detected inductively using two identical microstrip antennas separated by 5 mm. Each antenna was 50 μm wide and 2.5 mm long, fabricated on grounded Rogers 4003 substrates by photolithography and wet etching. The antennas were short-circuited at one end and connected to 50-Ω microstrip lines.

The $VO_2$ stripe was fabricated from a 650-nm-thick $VO_2$ film deposited on an amorphous quartz substrate by reactive DC magnetron sputtering in an $Ar/O_2$ atmosphere at 600°C. The film was patterned into 200-, 300-, and 500-μm-wide stripes by laser ablation using an ytterbium fiber laser (50 kHz repetition rate, 80 ns pulse duration, 60 μJ pulse energy) focused to a 50 μm spot with a galvanometric scanner. Large-area contacts were subsequently defined and coated with 2-μm-thick copper electrodes by thermal evaporation to enable current injection. The surface morphology and thickness of the $VO_2$ stripe were characterized by profilometry and atomic force microscopy. We also compared the IMT in patterned and continuous $VO_2$ films. For more details, see the Supplementary Materials.

### B. *Device characterization*

Spin-wave transport was characterized by measuring the $S_{11}$ and $S_{21}$ scattering parameters using a two-port VNA (Agilent N5222A or Rhode & Schwarz ZVB20). The microwave excitation power was fixed at –10 dBm to ensure linear spin-wave excitation. A DC current was applied to the $VO_2$ stripe and a 250-Ω series resistor using a source meter (Keithley 2410 or 2635A). The current was varied between 0 and 27 mA, corresponding to current densities $J$ up to $1.4\times10^4$ A cm$^{-2}$. An in-plane magnetic bias field was applied parallel to the microstrip antennas to saturate the YIG magnetization, establishing the Damon-Eshbach geometry ($k \perp H$). Measurements were performed over a field range of 58 – 196 mT. Temperature-dependent measurements were carried out using a temperature-controlled stage integrated into the microwave probe station.

Time-resolved $S_{21}$ measurements were performed by incorporating a MOSFET switchboard (High Voltage Experimental) into the measurement circuit (see Supplementary Materials for the circuit schematic). Voltage pulses (3.3 V) from an arbitrary waveform generator (AWG, Agilent 33220A) controlled the switchboard, enabling the injection of current pulses into the $VO_2$ stripe and the 250 Ω series resistor. The pulse duration and repetition rate were defined by the AWG, while the current amplitude was set by the source meter. The $S_{21}$ parameter was recorded by the VNA in time-domain mode at a fixed frequency.

### C. *Spin-wave transport and thermal modeling*

The insertion loss induced by the IMT in the $VO_2$ stripe was modeled using the approach of Ref. 43. The model computes the spin-wave dispersion for both the bare YIG film and the YIG/$VO_2$ bilayer and evaluates the resulting transmission characteristics. The experimentally observed downshift under applied current was reproduced by introducing a small reduction in $M_s$, attributed to moderate heating of the YIG film. The same framework was used to estimate spin-wave reflection as a function of $VO_2$ resistivity. Full details of the model, including input parameters and calculated dispersion relations, are provided in the Supplementary Materials.

Temperature distributions within the GGG substrate, YIG film, and $VO_2$ stripe under DC current were simulated using COMSOL Multiphysics. The simulation methodology and parameters are described in the Supplementary Materials.

## Acknowledgment

Work at Aalto University was supported by the Research Council of Finland (Grant Nos. 357211 and 359125). A.A.N. acknowledges financial support from the Magnus Ehrnrooth Foundation. Work at LUT was supported by the INERCOM platform.

# Supplementary Materials for

## Current control of spin-wave transport in YIG via the $VO_2$ insulator-to-metal transition

Aleksei A. Nikitin *et al.*
Corresponding authors:
Aleksei A. Nikitin, aleksei.a.nikitin@gmail.com
Sebastiaan van Dijken, sebastiaan.van.dijken@aalto.fi

### Surface topography of the $VO_2$ stripe and IMT characterization

The surface topography of the $VO_2$ stripe was characterized by atomic force microscopy (AFM) using a Park Systems XE7 operated in non-contact mode under ambient conditions. Figure S1A shows the surface topography over a 5×5 $\mu m^2$ area at the center of the stripe. Surface roughness was quantified following second-order image flattening, a procedure that does not appreciably affect features whose lateral extent is much smaller than the image size. The root mean square (rms) roughness of the 650-nm-thick $VO_2$ stripe is approximately 9 nm.

The geometry of the $VO_2$ stripe patterned by laser ablation was evaluated using a non-contact three-dimensional Sensofar S Neox optical profilometer (Fig. S1B). An optical micrograph of the patterned stripe, with the profilometer scan direction indicated by the white dashed line, is shown in the inset of Fig. S1B. Laser ablation produced a 1.5-µm-deep trench into the quartz substrate, effectively isolating the $VO_2$ stripe. The stripe surface is smooth, and the profilometer measurement confirms a uniform stripe thickness of approximately 650 nm. Minor imperfections are observed at the stripe edges; however, as these edges are not in contact with the YIG film in the magnonic device, they have no bearing on spin-wave transport.

The insulator-to-metal transition (IMT) was characterized before and after laser-ablation patterning of the $VO_2$ stripe. As shown in Fig. S1C, patterning results in a marginal reduction of the IMT temperature by ~1.5 K. Importantly, the resistivity contrast across the IMT is preserved upon patterning, confirming that the functional properties of the $VO_2$ stripe are not degraded by the fabrication process.

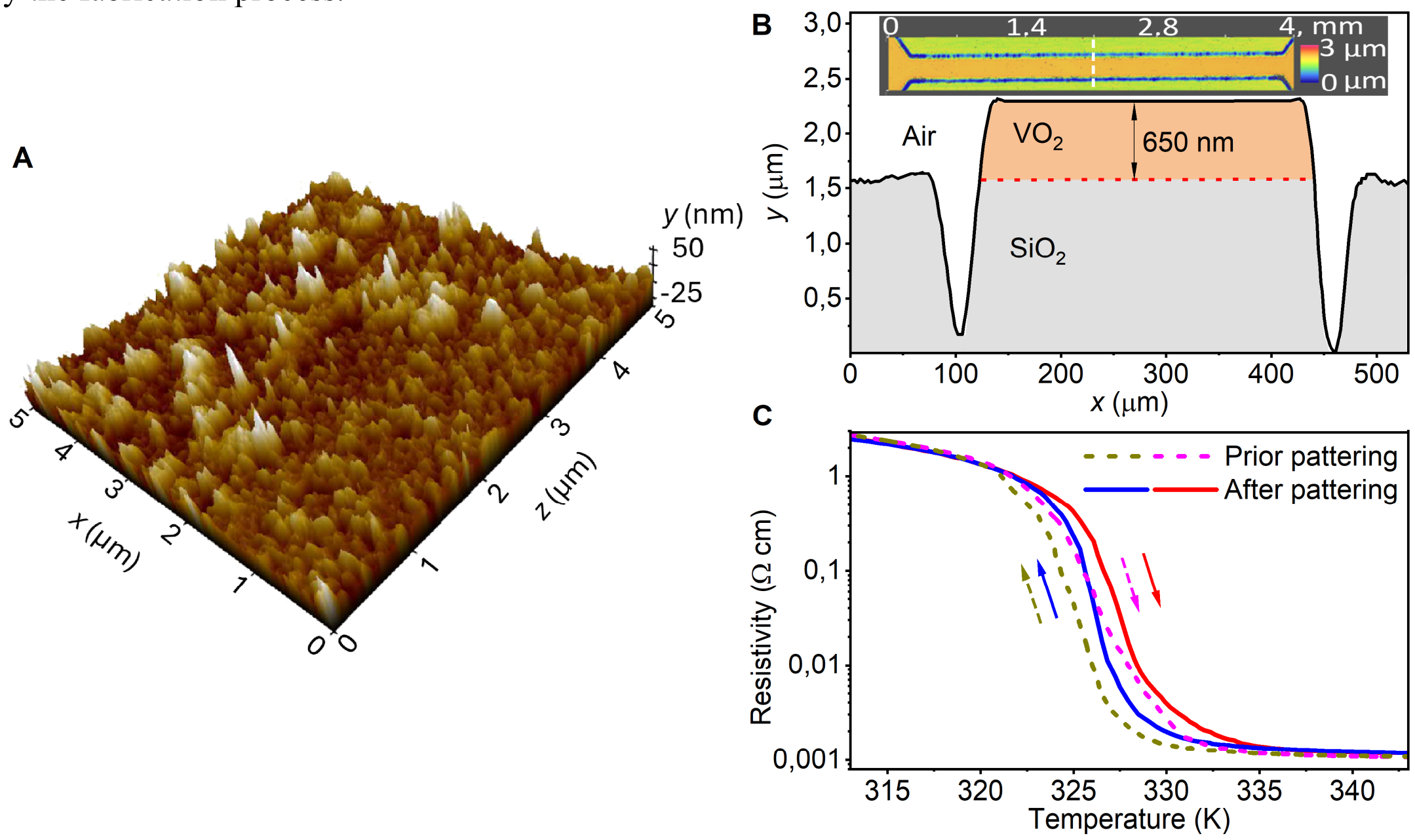

**Fig. S1. Surface topography and IMT of the $VO_2$ stripe. (A)** AFM image of the $VO_2$ stripe surface acquired over a 5×5 $\mu m^2$ area at the center of the stripe. Optical micrograph of the patterned $VO_2$ stripe (inset) and surface height profile measured along the stripe width by optical profilometry. **(C)** Temperature dependence of the electrical resistivity for the unpatterned $VO_2$ film and the laser-ablation-patterned $VO_2$ stripe.

## Theoretical model of spin-wave insertion loss

The insertion loss ($S_{21}$) of the magnonic device was calculated using the analytical model of Ref. 43. The YIG film was parameterized by thickness $t_1$ = 5.7 μm and saturation magnetization $M_{s,RT} = 1.39 \times 10^5$ A m$^{-1}$ at room temperature. The temperature dependence of the saturation magnetization was approximated by the linear relation $M_s(T) = M_{s,RT} - \eta(T - 298)$, with $\eta$ = 313 A K$^{-1}$ m$^{-1}$ (*33*), which is well justified over the relevant temperature range. The $VO_2$ stripe was parameterized by thickness $t_2$ = 650 nm and temperature-dependent resistivity $\rho_2(T)$, taken directly from the experimental data in Fig. 2A of the main text. The Damon-Eshbach configuration was established by applying a bias field perpendicular to the spin-wave propagation direction ($x$-axis shown in Fig. 1A).

The YIG film and adjacent $VO_2$ stripe were modeled as a rectangular waveguide boundary-value problem, accounting for magnetic losses in the YIG and the finite resistivity of both layers. The magnetic response of the YIG film is described by the complex permeability tensor:

$$\overset{\wedge}{\mu}_1 = \begin{pmatrix} \mu_d & j\mu_a & 0 \\ -j\mu_a & \mu_d & 0 \\ 0 & 0 & 1 \end{pmatrix}, \tag{1}$$

where $\mu_d = 1 + \frac{(\omega_H + j\alpha\omega)\omega_M}{(\omega_H + j\alpha\omega)^2 - \omega^2}$ and $\mu_a = \frac{\omega\omega_M}{(\omega_H + j\alpha\omega)^2 - \omega^2}$, with $\omega_H = \gamma\mu_0 H_0$, $\omega_M = \gamma\mu_0 M_s$, and the Gilbert damping parameter $\alpha = \frac{\gamma\Delta H}{\omega_H}$. Here, $\mu_0 = 4\pi \times 10^{-7}$ H m$^{-1}$ is the permeability of free space, $\gamma = 2.8 \times 10^{10}$ s$^{-1}$ T$^{-1}$ is the electron gyromagnetic ratio, $\omega$ is an angular frequency, and $\Delta H$ is the FMR linewidth. Magnetic loss was incorporated using the experimentally determined value $\Delta H$ = 31.83 A m$^{-1}$ at 6 GHz. This is justified by the well-established result that the linearized Landau-Lifshitz-Gilbert equation reduces to the linearized Landau-Lifshitz equation with a complex effective field (*64*). The electrical response of the $VO_2$ stripe is described by the complex permittivity:

$$\dot{\varepsilon}_2 = \varepsilon_2 - j\frac{1}{\varepsilon_0\omega\rho(T)}, \tag{2}$$

where $\varepsilon_0 = 8.85 \times 10^{-12}$ F m$^{-1}$ is the permittivity of free space and $\varepsilon_2$ = 1 is the relative permittivity of $VO_2$ in the insulating state.

The spin-wave dispersion relation for the YIG/$VO_2$ bilayer structure was obtained by solving the full set of Maxwell's equations subject to electromagnetic boundary conditions, treating the TE and TM modes as two independent systems, following Ref. 43. For the TE modes, the Helmholtz equation for the electric field component $E_z$ (derived from Maxwell's equations) in each layer $m$ takes the form:

$$\frac{\partial^2 E_z}{\partial y^2} + \frac{\partial^2 E_z}{\partial x^2} + \omega^2\mu_0\mu_m\varepsilon_0\varepsilon_m E_z = 0, \tag{3}$$

where $\mu_1 = (\mu_d^2 - \mu_a^2)/\mu_d$ for the YIG layer and $\mu_0$ = 1 for the $VO_2$ layer. The solutions of Eq. (3) are:

(a) in free space above and below the structure:

$$E_z = A_3 e^{-\kappa_0 y} e^{-jkx};\ E_z = A_0 e^{\kappa_0 y} e^{-jkx}, \tag{4}$$

(b) in the YIG and $VO_2$ layers:

$$E_z = (A_m \sin\kappa_m(y - D_m) + B_m \cos\kappa_m(y - D_m))e^{-jkx}, \tag{5}$$

where $\kappa_0 = \sqrt{k^2 - \omega^2\mu_0\varepsilon_0}$, $\kappa_1 = \sqrt{\omega^2\mu_0\varepsilon_0\varepsilon_1\mu_\perp - k^2}$, and $\kappa_2 = \sqrt{\omega^2\mu_0\varepsilon_0\varepsilon_2 - k^2}$ are the transverse wavenumbers in free space, the YIG film, and the $VO_2$ layer, respectively; $k$ is the spin-wave wavenumber; and $D_m = \sum_{s=1}^{2} t_s$ is the cumulative layer thickness. Note that $\kappa_1$ is imaginary within the spin-wave passband, $\omega_{FMR} = \sqrt{\omega_H(\omega_H + \omega_M)}$ to $\omega_H + \omega_M/2$, consistent with the surface-wave character of the Damon-Eshbach mode. The relationship between $H_x$ and $E_z$ follows from Maxwell's equations for each layer:

$$H_x = \frac{j(\mu\frac{\partial E_z}{\partial y} - k\alpha E_z)}{\omega\mu_0(\mu^2 - \alpha^2)};\ H_x = \frac{j}{\omega\mu_0}\frac{\partial E_z}{\partial y}. \tag{6}$$

Enforcing continuity of the tangential components $E_z$ and $H_x$ across all interfaces yields six homogeneous algebraic equations. Setting the determinant of the resulting coefficient matrix to zero gives the exact dispersion relation:

$$\begin{vmatrix} \kappa_0 & -\dfrac{k\mu_a \tan(\kappa_1 t_1) + \mu_d \kappa_1}{\mu_1 \mu_d} & \dfrac{k\mu_a - \kappa_1 \mu_d \tan(\kappa_1 t_1)}{\mu_1 \mu_d} & 0 & 0 & 0 \\ 1 & \tan(\kappa_1 t_1) & -1 & 0 & 0 & 0 \\ 0 & -\dfrac{\kappa_1}{\mu_1 \mu_d} & \dfrac{k\mu_a}{\mu_1 \mu_d} & \kappa_2 & \kappa_2 \tan(\kappa_2 t_2) & 0 \\ 0 & 0 & -1 & -\tan(\kappa_2 t_2) & 1 & 0 \\ 0 & 0 & 0 & -\kappa_2 & 0 & -\kappa_0 \\ 0 & 0 & 0 & 0 & -1 & 1 \end{vmatrix} = 0, \tag{7}$$

The vanishing of this determinant yields the complex-valued dispersion relation $\dot{\omega}(k) = \omega(k) + j\omega_r(k)$, whose real part determines the spin-wave group velocity $v_g = \delta\omega/\delta k$ and whose imaginary part characterizes the spatial damping decrement $\alpha = 2\pi\omega_r/v_g$.

To further substantiate the IMT effect on spin-wave dynamics, we performed additional calculations at three applied magnetic fields: 58 mT (low-field limit in Fig. 5B), 144 mT (corresponding to Fig. 3B and the black square in Fig. 5B), and 178 mT (high-field limit in Fig. 5B). Figure S2A shows the calculated spin-wave dispersion relations in YIG/$VO_2$ as a function of applied field for the insulating state of $VO_2$, corresponding to a resistivity of $\rho_3 = 3.7$ Ω·cm ($J = 0$ A cm$^{-2}$). A slight reduction in the slope of the dispersion curves is observed with increasing field, consistent with a decrease in spin-wave group velocity. Notably, the dispersions at $J = 0$ A cm$^{-2}$ are nearly indistinguishable from those of bare YIG, confirming that the insulating $VO_2$ stripe has negligible influence on spin-wave propagation. The situation changes markedly when $VO_2$ is driven into the metallic state, either partially at $\rho_3 = 2.0$ mΩ·cm ($J = 1.1 \times 10^4$ A cm$^{-2}$, blue lines) or more completely at $\rho_3 = 1.2$ mΩ·cm ($J = 1.4 \times 10^4$ A cm$^{-2}$, red lines). The transition to a conducting $VO_2$ state substantially modifies the spin-wave dispersion relation, which now exhibits a pronounced dependence on both resistivity and applied magnetic field. These dispersion changes directly govern the spatial damping decrement $\alpha$, which is determined by both the spin-wave group velocity and the imaginary part of the complex-valued wave vector. As shown in Supplementary Fig. S2B, the low spin-wave attenuation characteristic of the insulating state ($J = 0$ A cm$^{-2}$) gives way to strong transmission suppression upon IMT-induced metallization of the $VO_2$ stripe at $J = 1.1 \times 10^4$ A cm$^{-2}$ and $J = 1.4 \times 10^4$ A cm$^{-2}$. Importantly, the magnitude of this suppression is sensitively dependent on the applied magnetic field, with stronger damping observed at higher fields, in quantitative agreement with the experimental data on spin-wave attenuation as a function of magnetic field presented in Fig. 5B of the main text.

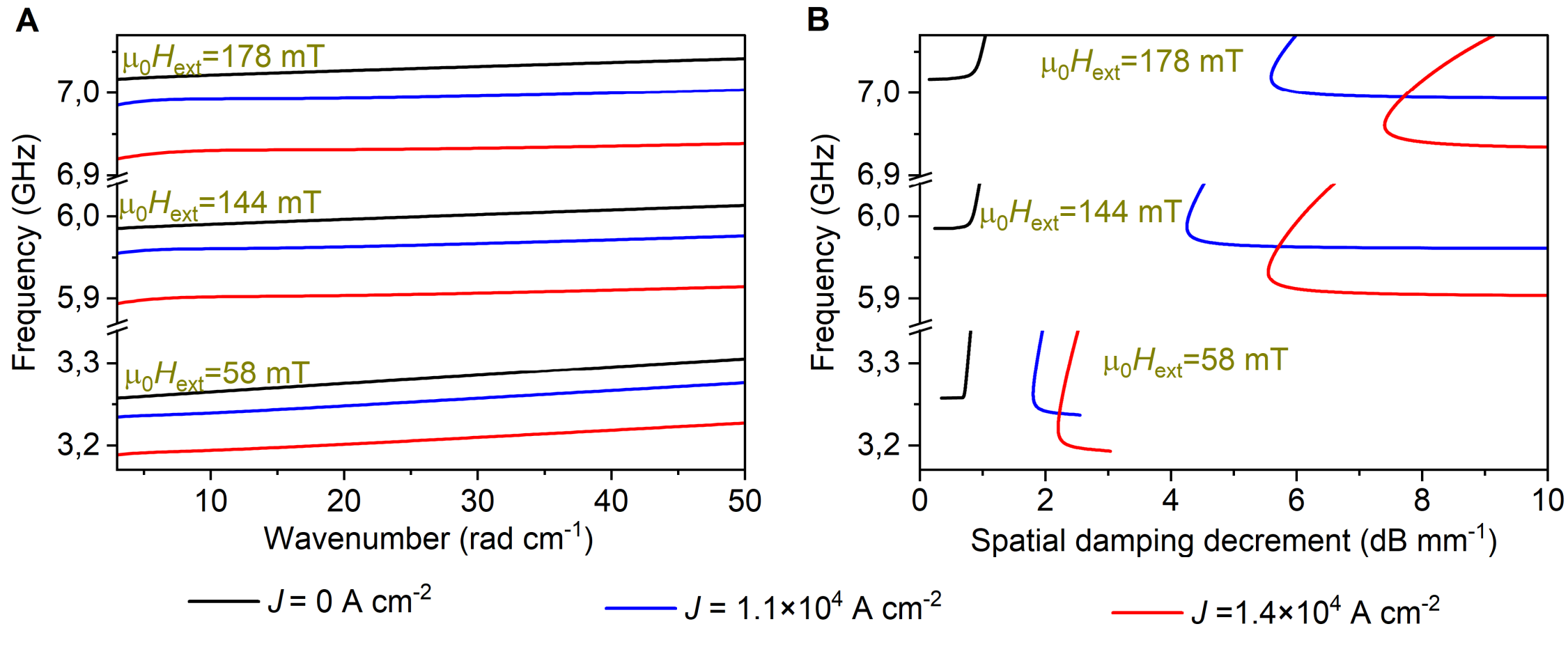

**Fig. S2.Magnetic-field dependence of spin-wave dispersion and spatial damping in YIG/$VO_2$. (A)** Calculated spin-wave dispersion relations at three applied magnetic fields (58 mT, 144 mT, and 178 mT) for current densities $J = 0$ A cm$^{-2}$ (black solid lines), $J = 1.1 \times 10^4$ A cm$^{-2}$ (blue solid lines), and $J = 1.4 \times 10^4$ A cm$^{-2}$ (red solid lines), corresponding to $VO_2$ resistivities of approximately 3.7 Ω·cm, 2.0 mΩ·cm, and 1.2 mΩ·cm, respectively. **(B)** Spatial damping decrement as a function of applied magnetic field for the same three current densities.

The same theoretical framework was employed to calculate the spin-wave power-reflection coefficient as a function of $VO_2$ resistivity, using an integral-equation formulation of the scattering problem. Extending the phenomenological approach of Ref. *65*, which treats reflection of Damon-Eshbach spin waves from a perfectly conducting stripe, the power-reflection coefficient for a $VO_2$ stripe of width $w_2 = 300$ µm on a YIG film of thickness $t_1$ takes the form:

$$\Gamma = i\pi w_2^2(k - k^*)\frac{\sinh(k^* t_1)}{\sinh(k t_1)}\frac{\mu_d - j\mu_a}{1 - j\mu_a + \mu_d}\frac{0.25 + \beta^2}{\cosh(\pi\beta)}\frac{1 - e^{-2kt_1}}{4t_1}\sqrt{\frac{(1+\mu_d)^2 + \mu_a^2}{\mu_d^2 + \mu_a^2}}, \quad (8)$$

where $k$ and $k^*$ are the real parts of the wavenumbers for spin waves propagating in the forward and backward directions, respectively, and $\beta = \frac{1}{2\pi} ln\left[\frac{1-j\mu_a+\mu_d}{1+j\mu_a+\mu_d}\frac{\mu_d+j\mu_a}{\mu_d-j\mu_a}\right]$ with $\mu_a$ and $\mu_b$ elements of the magnetic permeability tensor (see Eq. 1).

The calculated power-reflection coefficient as a function of excitation frequency and $VO_2$ resistivity is shown in Fig. S3. In the insulating state of $VO_2$, the reflection coefficient remains below 0.03 across the entire experimental frequency range, confirming that the insulating $VO_2$ stripe does not appreciably scatter propagating spin waves. Upon transition to the metallic state at $J = 1.4 \times 10^4$ A cm$^{-2}$, the coefficient increases from 0.01 to a maximum of 0.1, implying that no more than 10% of the incident spin-wave power is back-reflected at any frequency. Spin-wave reflections therefore remain a minor perturbation to transport under the conditions studied, consistent with the experimental $S_{11}$ spectra in Fig. 3A of the main text, which exhibits only weak oscillations above $f_{FMR}$ and no substantial changes in the reflection baseline upon current injection.

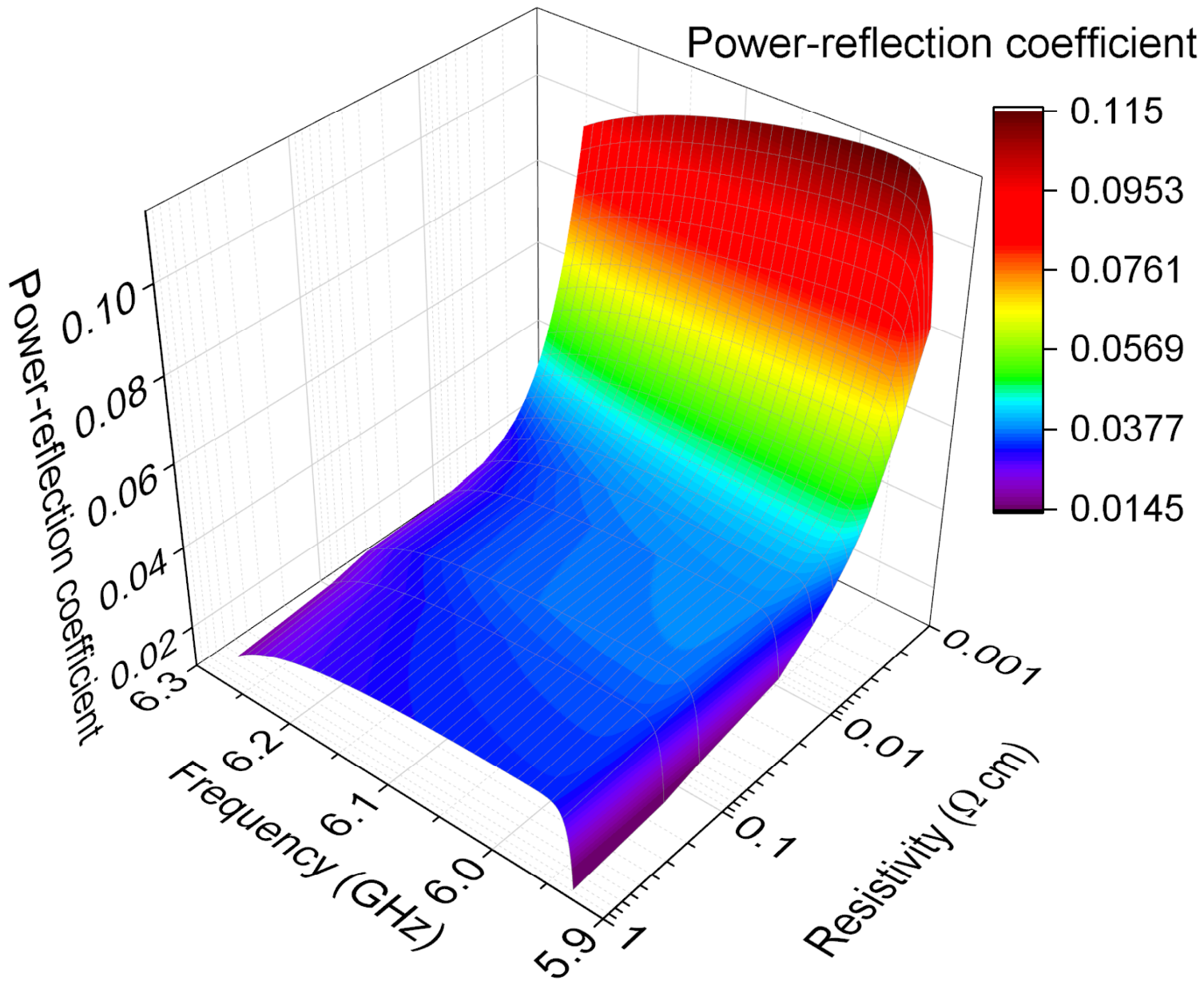

**Fig. S3. Calculated power-reflection coefficient for Damon–Eshbach spin waves scattered by the $VO_2$ stripe.** Three-dimensional map of the power-reflection coefficient for Damon-Eshbach spin waves propagating in a 5.7-µm-thick YIG film, shown as a function of excitation frequency and $VO_2$ resistivity, spanning the full range of the IMT.

To complete the modelling, we assessed the power transmission coefficient of the full device. The YIG waveguide structure, with a total length of $L = 5$ mm, was partitioned into three regions. Region I, of length $l_I = 2.35$ mm, corresponds to the uncovered YIG film between the exciting antenna and the left edge of the $VO_2$ stripe. Region II, of length $l_{II} = 300$ µm, corresponds to part of the YIG film that is contacted by the $VO_2$ stripe. Region III, with the same length as region I ($l_I = 2.35$ mm), spans the uncovered YIG film between the right edge of the $VO_2$ stripe and the detecting antenna. The total power transmission coefficient is then given by:

$$S_{21}(\omega) = 20\log\left(\frac{(\Gamma^2 - 1)\, e^{(-ik_{II}+\alpha_{II})l_{II}+(ik_I-\alpha_I)2l_I}}{\Gamma^2 - e^{2(-ik_{II}+\alpha_{II})l_{II}}}\right), \quad (9)$$

where $k_I$ and $k_{II}$ are the spin-wave wavenumbers in the uncovered YIG film and the YIG/$VO_2$ bilayer, respectively, and $\alpha_I$ and $\alpha_{II}$ are the corresponding spatial damping decrements in each section of the magnonic device.

## Oersted field generated by the applied current

As discussed in the main text, application of a DC current to the $VO_2$ stripe produces a pronounced downshift of the FMR frequency visible in Fig. 3. This behavior cannot be attributed to a current-induced Oersted field, since the FMR frequency shift is independent of the current direction in our experiments. To corroborate this conclusion, we calculated the magnitude of the Oersted field generated around the $VO_2$ stripe carrying a DC current. Following the field analysis of Ref. *66*, a current $I$ flowing uniformly through a stripe with cross-sectional coordinates ($x$, $y$) produces, at the observation point ($X$, $Y$), a magnetic field component perpendicular to the stripe given by:

$$B(X,Y) = \frac{\mu_0 I}{8\pi ab} \int_{-a}^{a} \int_{-b}^{b} \frac{(y-Y)}{(x-X)^2 + (y-Y)^2} \partial x \partial y \qquad (10)$$

where the coordinate origin is placed at the center of the stripe cross-section. Here, $a = t_2/2 = 325$ nm and $b = w/2 = 150$ µm define the half-thickness and half-width of the $VO_2$ stripe, respectively; $X = 0$ and $Y = t_1/2 = 2.85$ µm are the coordinates of the center of the YIG film directly above the stripe; and $I = 27$ mA is the maximum current applied in the experiments. The calculated in-plane Oersted field at this position is 0.016 mT, which is negligibly small compared to the applied bias field of 144 mT and has no measurable effect on spin-wave transport in the device.

## Insertion loss of the bare YIG film and YIG/$VO_2$ device without applied current

The presence of the insulating $VO_2$ stripe does not measurably affect spin-wave transmission in the YIG film, consistent with its negligible influence on the spin-wave dispersion in the insulating state. To demonstrate this, the insertion loss ($S_{21}$) was measured at room temperature for a bare YIG film and for the same film in contact with the $VO_2$ stripe, using two microstrip antennas separated by 5 mm. The results are shown in Fig. S4 as red dashed and black solid lines, respectively.

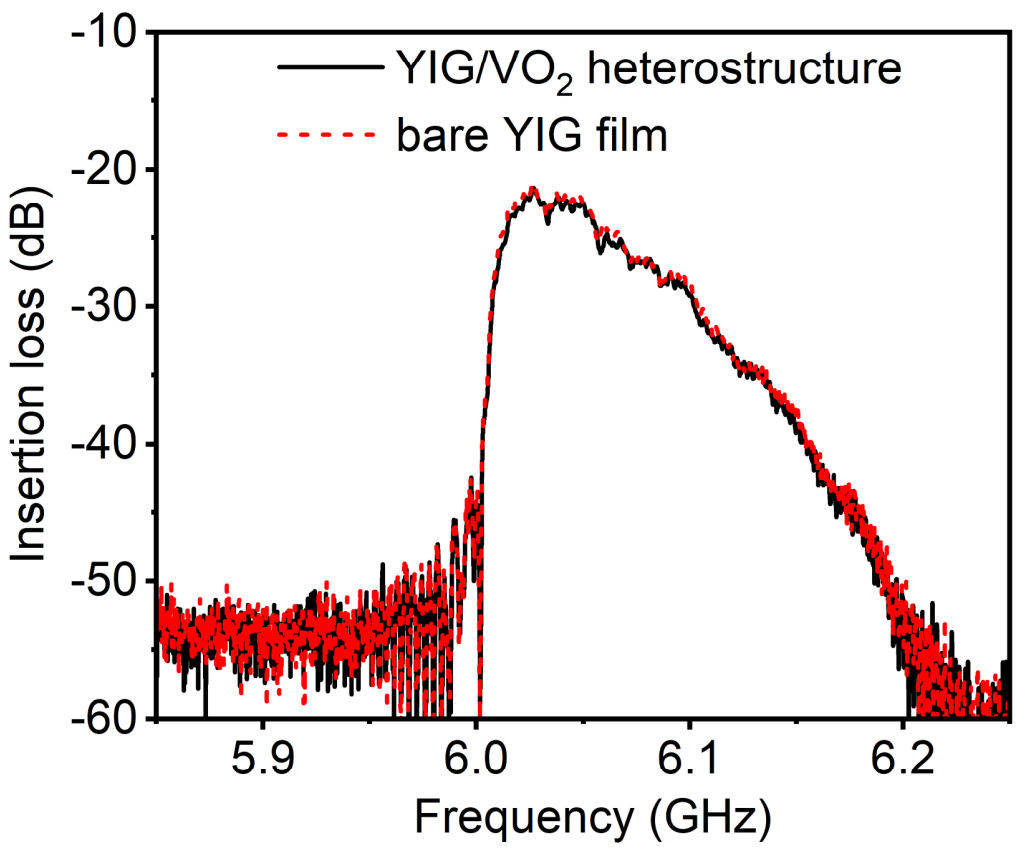

**Fig. S4. Spin-wave insertion loss of the bare YIG film and YIG/$VO_2$ device.** Real parts of the $S_{21}$ spectra measured at room temperature on a bare YIG film (red dashed line) and the same film in contact with a 300-µm-wide insulating $VO_2$ stripe (black solid line). No current is applied to the stripe. The two spectra are indistinguishable, confirming that the insulating $VO_2$ stripe has no measurable effect on spin-wave transmission.

## Magnetic-field dependence of spin-wave transmission spectra

To establish the magnetic-field dependence of the spin-wave suppression shown in Fig. 5B of the main text, we recorded $S_{21}$ spectra at a fixed current density of $J = 1.4 \times 10^4$ A cm$^{-2}$ under varying applied magnetic fields. A representative selection of spectra is presented in Fig. S5.

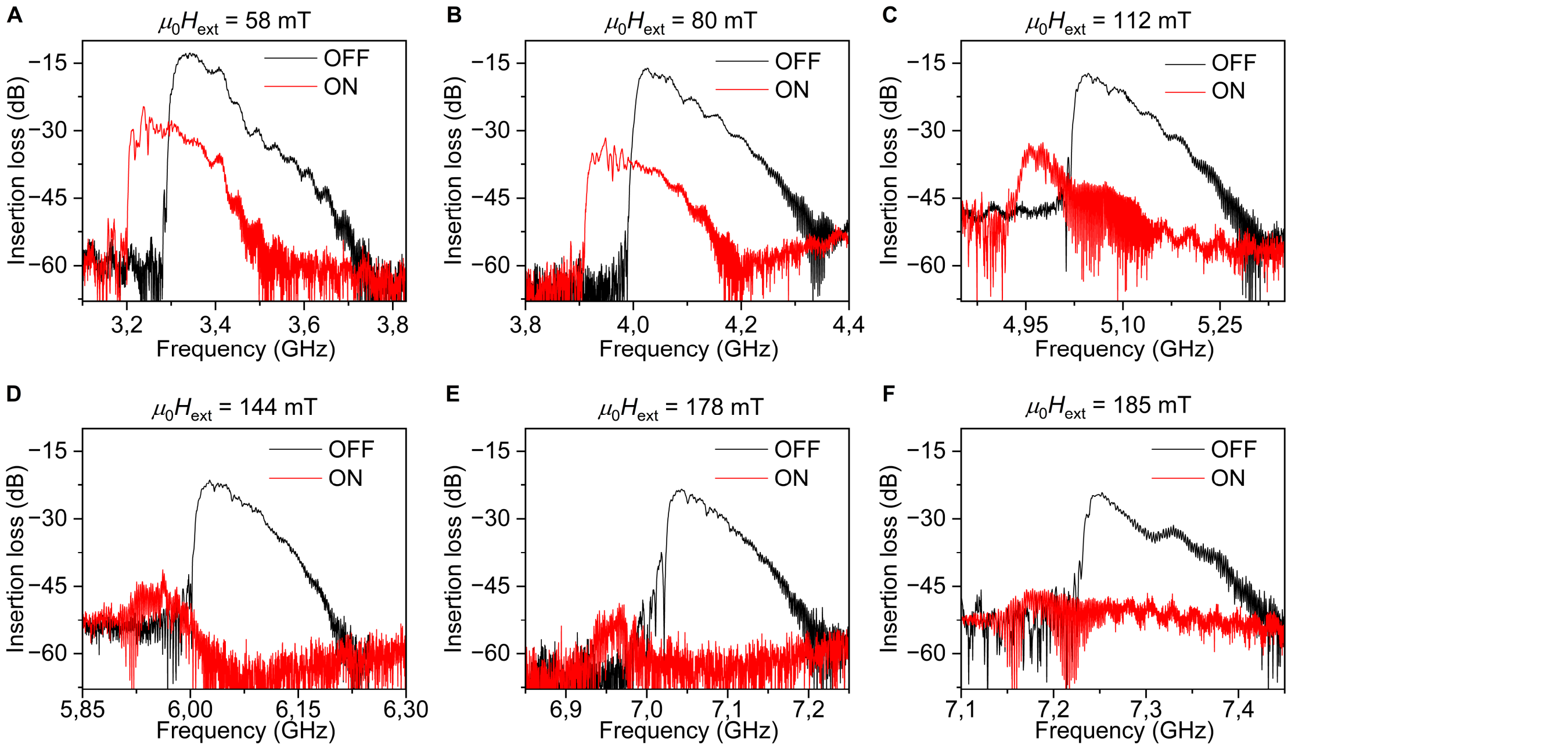

**Fig. S5. Magnetic-field dependence of spin-wave transmission spectra. (A-F)** Real parts of the $S_{21}$ spectra recorded on a YIG film with a 300-µm-wide $VO_2$ stripe at six representative bias fields. Black curves correspond to measurements with the current off ($J = 0$ A cm$^{-2}$); red curves correspond to measurements with the current on ($J = 1.4 \times 10^4$ A cm$^{-2}$). With increasing magnetic bias field, the spin-wave transmission spectra shift to higher frequencies, consistent with the Kittel-like dispersion of the FMR frequency $f_{FMR}$, and the current-induced suppression of the spin-wave signal increases.

## Setup for time-resolved spin-wave transmission measurements

Time-resolved $S_{21}$ measurements were performed by incorporating a MOSFET switchboard into the measurement circuit. Voltage pulses (3.3 V) from an arbitrary waveform generator (AWG) controlled the switchboard, enabling the injection of current pulses into the $VO_2$ stripe and the 250 Ω series resistor ($R_E$). A large load resistor ($R_L$ = 100 kΩ) placed in parallel ensured that the source meter operated in constant-current mode when the switchboard was open. The pulse duration and repetition rate were set by the AWG. The $S_{21}$ scattering parameter was recorded by the VNA in time-domain mode at a fixed excitation frequency. Figure S6 shows an optical micrograph of the magnonic device together with a schematic of the measurement configuration.

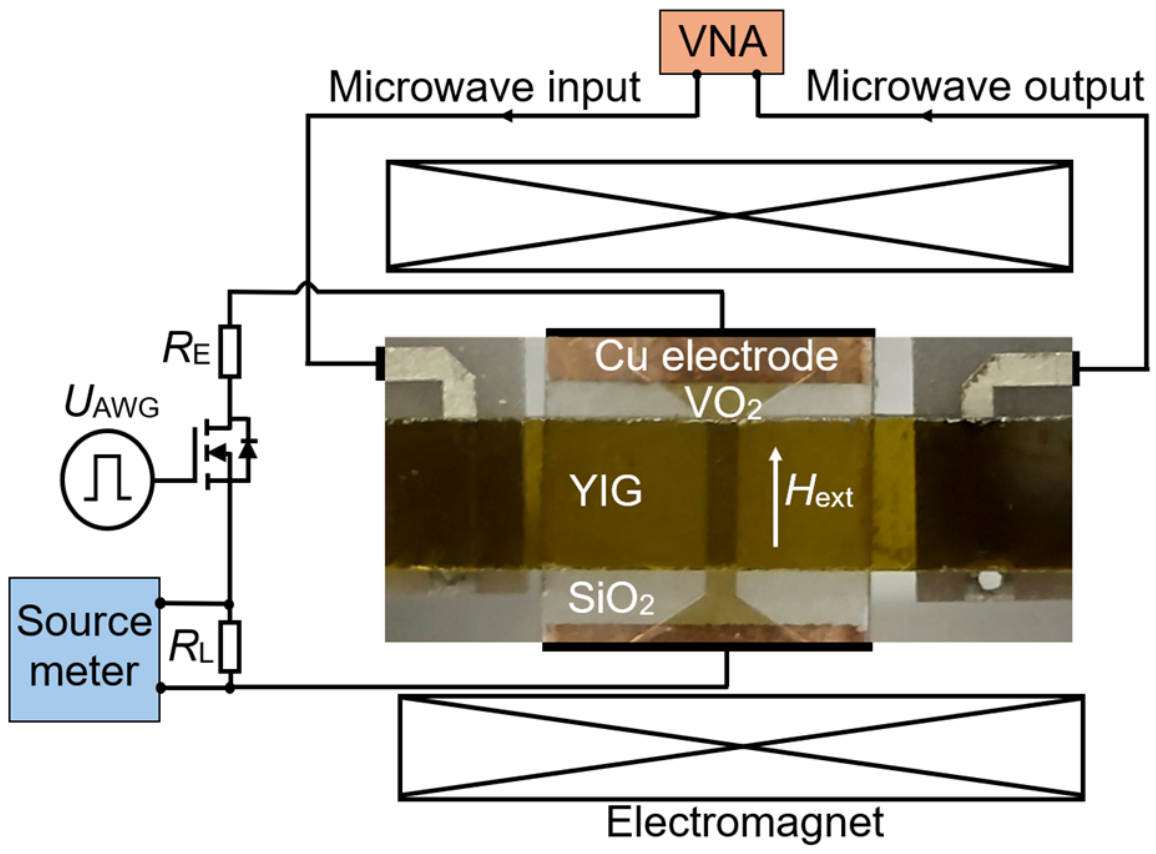

**Fig. S6. Time-resolved measurement configuration.** Optical micrograph and circuit schematic of the magnonic device and measurement setup. Spin waves are excited and detected by the microstrip antennas via a VNA. The resistivity of the $VO_2$ stripe is modulated by trains of current pulses by toggling the MOSFET switchboard with voltage pulses from the AWG. When the switchboard is open (off state), current from the source meter is routed through the load resistor ($R_L$ = 100 kΩ). When the switchboard is closed (on state), current flows predominantly through the $VO_2$ stripe and the series resistor ($R_E$ = 250 Ω). The duration and repetition rate of the current pulses are defined by the AWG voltage waveform.

## Time-resolved measurements with varying current-pulse duration

The temporal response of the $S_{21}$ signal to a current-pulse train was recorded at a base temperature of $T = 318$ K and a current density of $J = 1.0 \times 10^3$ A cm$^{-2}$ for a fixed repetition rate of 10 Hz and three pulse durations. The single-pulse response is presented in Fig. 6A of the main text; the multi-pulse response for varying pulse duration is shown in Fig. S7.

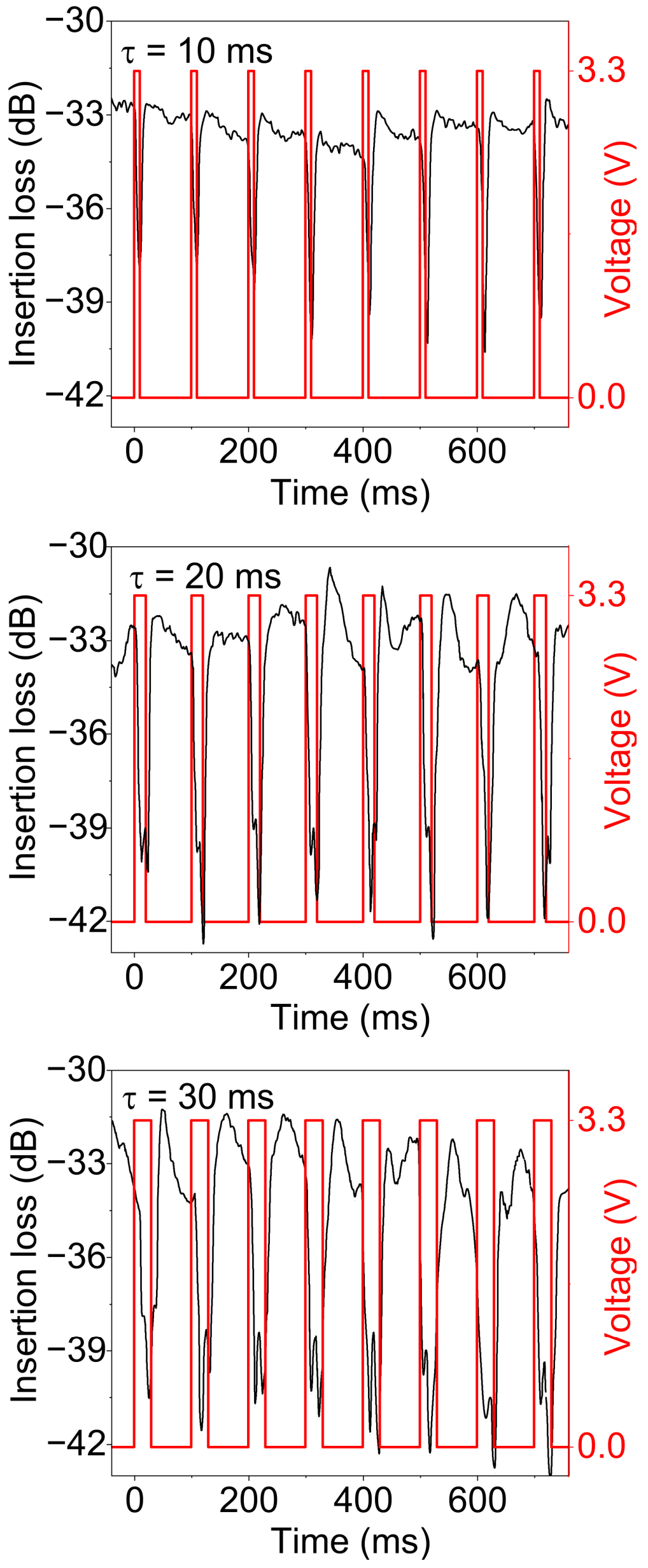

**Fig. S7. Time-resolved $S_{21}$ response to current-pulse trains with varying pulse durations.** Temporal evolution of the $S_{21}$ signal recorded under a current-pulse train with pulse durations of $\tau = 10$, 20, and 30 ms. The current density is $J = 1.0 \times 10^3$ A cm$^{-2}$, the pulse repetition rate is 10 Hz, the base temperature is 318 K, the spin-wave excitation frequency is 7.49 GHz, and the applied magnetic field is 196 mT.

## Current-voltage characteristic of the current-induced IMT in $VO_2$ at 318 K

Most experiments utilizing DC currents were performed at a base temperature of 298 K. The corresponding current-voltage characteristic of the $VO_2$ stripe is shown in Fig. 2C of the main text. Time-resolved measurements employing current pulses were performed at 318 K, closer to the IMT temperature, where a lower threshold current is required to drive the IMT. The current-voltage characteristic of the $VO_2$ stripe at this temperature is presented in Fig. S8.

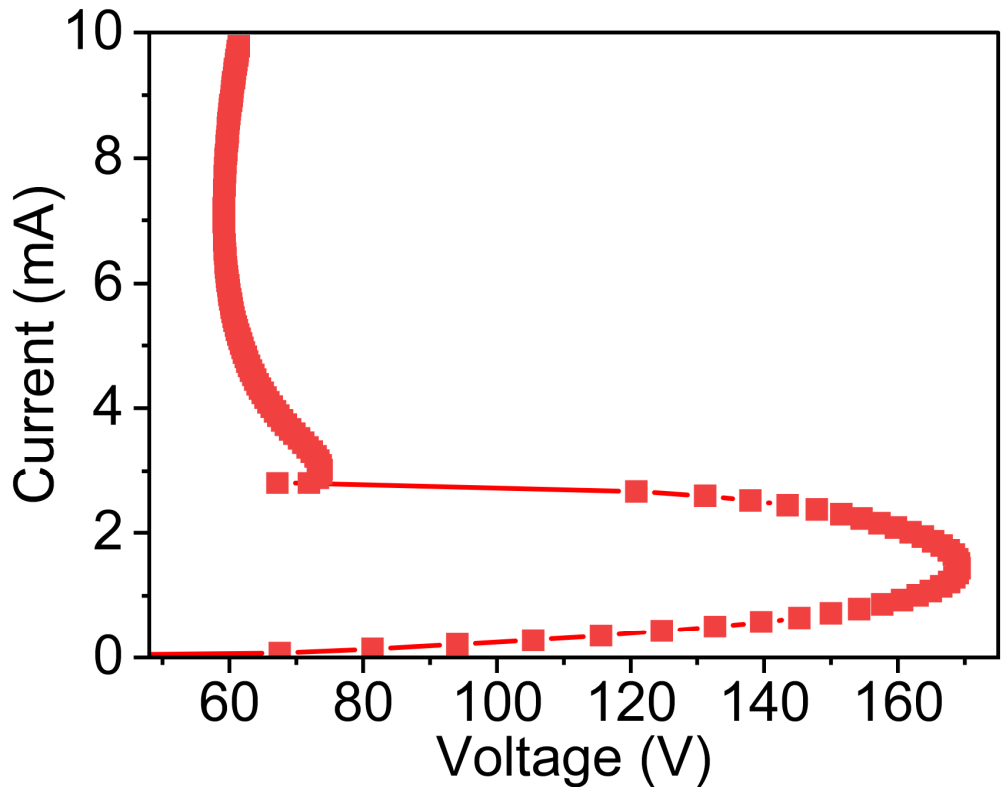

**Fig. S8. Current-voltage characteristic of the current-driven IMT in $VO_2$ at 318 K.** Voltage across the $VO_2$ stripe as a function of applied current at 318 K. Compared with the 298 K measurement shown in Fig. 2C of the main text, the IMT occurs more abruptly and at a lower threshold current, consistent with the proximity of 318 K to the intrinsic IMT transition temperature of $VO_2$.

## Performance scaling with $VO_2$ stripe width and YIG film thickness

To assess the scaling of current-controlled spin-wave transmission in reduced device geometries, we employed the theoretical model outlined in Supplementary Note II. Figure S9A shows how the suppression of spin-wave transmission at $J = 1.1 \times 10^4$ A cm$^{-2}$ scales with $VO_2$ stripe width for four YIG film thicknesses. The linear scaling is in excellent agreement with experimental data, as confirmed by $S_{21}$ spectra measured on devices with $VO_2$ stripe widths of 200, 300, and 500 µm (Fig. S9A, insets). This linear dependence confirms that the transmission suppression originates from a current-induced modification of the spin-wave group velocity in the region of the YIG film that is in contact with the $VO_2$ stripe, rather than from bulk losses. Figure S9B shows the dependence of the transmission suppression on YIG film thickness. In contrast to the stripe-width dependence, this relationship is non-linear: reducing the film thickness generally decreases the degree of spin-wave suppression, with the rate of change depending on the thickness regime.

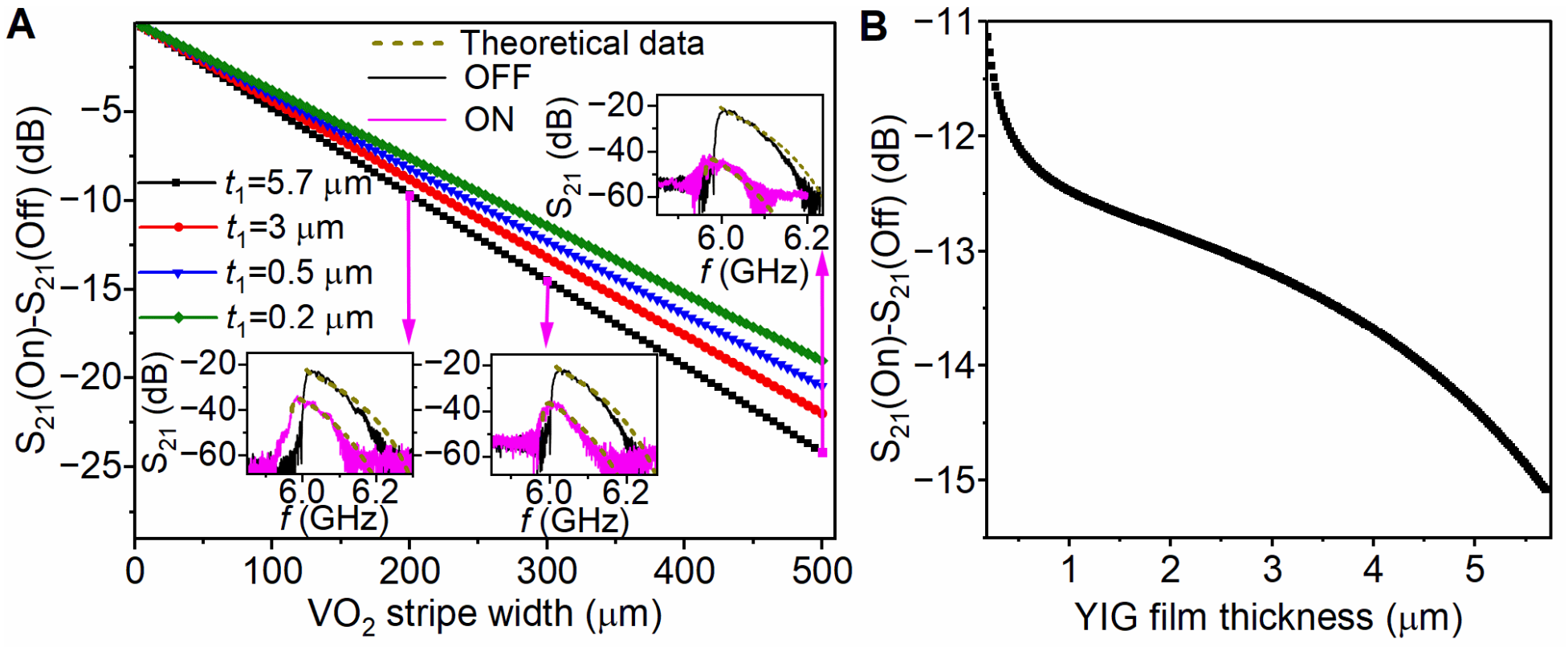

**Fig. S9. Scaling of current-induced spin-wave suppression with $VO_2$ stripe width and YIG film thickness.** (**A**) Calculated suppression of spin-wave transmission at $1.006 f_{FMR}$ under $J = 1.1 \times 10^4$ A cm$^{-2}$ as a function of $VO_2$ stripe width, for four YIG film thicknesses. Insets show experimental $S_{21}$ spectra recorded on devices with $VO_2$ stripe widths of 200, 300, and 500 µm. (**B**) Calculated suppression of spin-wave transmission at $1.006 f_{FMR}$ as a function of YIG film thickness, for a $VO_2$ stripe width of 300 µm and $J = 1.1 \times 10^4$ A cm$^{-2}$. Reducing the YIG film thickness generally decreases the achievable spin-wave suppression.

## COMSOL Multiphysics simulations of temperature distributions

Three-dimensional finite element simulations of the temperature distribution in the GGG substrate, YIG film, and $VO_2$ stripe under applied DC current were performed in COMSOL Multiphysics. The thermal properties of the GGG and YIG layers and their temperature dependencies were taken from the COMSOL Multiphysics materials library. The thermal parameters of the $VO_2$ stripe required separate treatment, as described below.

The heat capacity $C_{VO}$ was calculated from the $VO_2$ mass density of 4340 kg·m$^{-3}$ (*67*) and the specific heat capacity per unit volume (*68*). The total thermal response of the $VO_2$ stripe reflects contributions from both electrical conduction (derived from resistivity data in Fig. 2 of the main text) and phonon thermal conduction $\lambda_{VO}$. The effect of the IMT temperature $T_{IMT}$ on $\lambda_{VO}$ was neglected in the simulations, on the grounds that the structural transformation of the $VO_2$ lattice across $T_{IMT}$ involves only small displacements of the V atoms (*69*). In the temperature range of interest, phonon thermal conduction in $VO_2$ is governed by a combination of impurity scattering, phonon-phonon interactions, and Umklapp processes, consistent with a Debye temperature of $\theta_D$ = 750 K (*68*). The temperature dependence of $\lambda_{VO}$ was described by the empirical relation reported in Ref. *70*:

$$\lambda = a \cdot \left(\frac{T}{\theta_D}\right)^n \cdot \exp\left(\frac{\theta_D}{b \cdot T}\right), \tag{11}$$

where $a$, $n$, and $b$ are phenomenological parameters with values 0.89, 1, and 1.01, respectively, determined for thin $VO_2$ layers (*71, 72*). The thermal diffusivity $\kappa_{VO}$ was obtained from the relation $\kappa_{VO} = \lambda_{VO}/C_{VO}$, and the average coefficient of thermal expansion $\alpha_{VO}$ was approximated by piecewise linear fits to the data of Ref. *73*. All $VO_2$ thermal parameters used in the simulations are compiled in Table S1.

**Table S1.** Thermal and electrical parameters of the $VO_2$ stripe used in the three-dimensional finite element simulations. Values are listed as a function of temperature $T$ for the heat capacity $C_{VO}$, thermal conductivity $\lambda_{VO}$, thermal diffusivity $\kappa_{VO}$, coefficient of thermal expansion $\alpha_{VO}$, and electrical conductivity $\sigma_{VO}$ (heating semi-cycle).

| $T$ (K) | $C_{VO}$ (J kg$^{-1}$ K$^{-1}$) | $\lambda_{VO}$ (W m$^{-1}$ K$^{-1}$) | $\kappa_{VO}$ (10$^{-6}$ m$^2$ s$^{-1}$) | $\alpha_{VO}$ (K$^{-1}$) | $\sigma_{VO}$ (S cm$^{-1}$) |
|---|---|---|---|---|---|
| 280 | 638.92 | 4.713 | 1.7 | 5.7 | 29.35 |
| 284 | 645.13 | 4.605 | 1.645 | 5.7 | 29.86 |
| 288 | 650.43 | 4.503 | 1.595 | 5.7 | 30.54 |
| 292 | 654.28 | 4.407 | 1.552 | 5.7 | 31.31 |
| 296 | 658.48 | 4.317 | 1.51 | 5.7 | 32.07 |
| 300 | 662.67 | 4.231 | 1.471 | 5.7 | 33.17 |
| 304 | 667.22 | 4.15 | 1.433 | 5.7 | 34.56 |
| 308 | 671.77 | 4.073 | 1.397 | 5.7 | 36.51 |
| 312 | 677.36 | 4.001 | 1.361 | 5.7 | 38.88 |
| 316 | 682.61 | 3.932 | 1.327 | 5.7 | 49.99 |
| 320 | 687.85 | 3.866 | 1.295 | 5.7 | 75.17 |
| 324 | 692.75 | 3.804 | 1.265 | 5.7 | 169.3 |
| 328 | 697.64 | 3.745 | 1.237 | 13.35 | 8782 |
| 332 | 701.84 | 3.688 | 1.211 | 13.35 | 48562 |
| 336 | 706.39 | 3.635 | 1.186 | 13.35 | 76835 |
| 340 | 710.23 | 3.584 | 1.163 | 13.35 | 83395 |